\documentstyle[12pt]{article}
\title{Non-Parallel Electric and Magnetic Fields in a Gravitational Background, Stationary Gravitational Waves and Gravitons}
\author{*Carlos Pinheiro - **J.A. Helay${\ddot{\mbox{e}}}$l Neto - ***Gilmar S. Dias  \\
(*Departamento de F\'{\i}sica, CCE, UFES. **Centro Brasileiro\\ de Pesquisas  F\'{\i}sicas, CBPF. Univesidade Cat\'olica de Petr\'o\_ \\polis, UCP. ***Departamento de F\'{\i}sica, CCE, \\
UFES. Escola T\'ecnica Federal ETFES)}
\date{ }

\newcommand{\gt}{\tilde{g}}
\newcommand{\guv}{g_{\mu \nu}}

\newcommand{\fuv}{F_{\mu \nu}}
\newcommand{\Fuv}{F^{\mu \nu}}

\newcommand{\Flv}{F^{\lambda \nu}}

\newcommand{\Ful}{F^{\mu \lambda}}

\newcommand{\fvb}{F_{\nu \beta}}

\newcommand{\fbu}{F_{\beta \mu}}

\newcommand{\Fub}{F^{\mu \beta}}

\newcommand{\au}{A_{\mu}}

\newcommand{\av}{A_{\nu}}

\newcommand{\dv}{\partial_\nu}

\newcommand{\du}{\partial_\mu}

\newcommand{\deu}{{\cal D}_\mu}

\newcommand{\dev}{{\cal D}_\nu}

\newcommand{\deb}{{\cal D}_\beta}

\newcommand{\Gvul}{\Gamma^{\nu}_{\mu\lambda}}

\newcommand{\Gbbl}{\Gamma^{\beta}_{\beta\lambda}}

\newcommand{\dgt}{\frac{\partial{g}}{\partial{t}}}
\newcommand{\dft}{\frac{\partial{f}}{\partial{t}}}

\newcommand{\ddft}{\frac{\partial^2{f}}{\partial{t^2}}}

\newcommand{\ddBt}{\frac{\partial^2\vec{B}}{\partial{t^2}}}
\newcommand{\ddEt}{\frac{\partial^2\vec{E}}{\partial{t^2}}}
\newcommand{\dBt}{\frac{\partial\vec{B}}{\partial{t}}}
\newcommand{\dEt}{\frac{\partial\vec{E}}{\partial{t}}}

\newcommand{\vE}{$\vec{E}$}
\newcommand{\vB}{$\vec{B}$}

\newcommand{\na}{\vec\nabla}

\begin{document}
\maketitle
\begin{abstract}
The existence of an electromagnectic field with parallel electric and magnetic components is readdressed in the presence of a gravitational field. A non-parallel solution is shown to exist. Next, we analyse  the possibility of finding  stationary gravitational waves in de nature. Finaly, We construct a D=4 effecttive quantum gravity model. Tree-level unitarity is verified.
\end{abstract}	
\newpage

%%%%%%%Equacao numero (1)
\section{Electric and Magnetic Field in a Gravitational Background}
Based on a series of papers by Brownstein \cite{K R Browns} and Salingaros \cite{Saling1}, we readdress here the possibility of the existence of an electromagnetic field whose electric and magnetic components are parallel in the presence of a gravitational field. The coupling between the electromagnetic sector and the gravitational background is accomplished by means of the action.

\begin{equation}
\label{acao}
{\cal S}=\int\sqrt{-\tilde{g}}(-\frac{1}{4}F_{\mu\nu}F^{\mu\nu})d^4x,
\end{equation}
where \[\gt=det(\guv),\]
and
\[\fuv=\du\av-\dv\au.\]
From the above action, the following field-equations follow:
%%%%%%% equacao numero (2)
\begin{equation}
\label{equacao de campo 1}
\deu\Fuv=\du\Fuv+\Gbbl\Flv+\Gvul\Ful=J^\nu
\end{equation}

\begin{equation}
\label{equacao de campo 2}
\deu\fvb+\dev\fbu+\deb\fuv=0.
\end{equation}
Choosing the background to be described by the F.R.W metric,

\begin{equation}
\label{metrica}
d{\mbox{S}}^2=dt^2-a^2(t)\biggl[\frac{dr^2}{1-Ar^2}+r^2d\theta^2+r^2\sin^2\theta{d}\phi^2\biggr],
\end{equation}
the Maxwell equations in the absence of electromagnetic sources read as below:

\begin{equation}
\label{equacoes maxwell}
\vec\nabla\cdot\vec{E}=g\vec\nabla{f}\cdot\vec{E}
\end{equation}
\[\vec\nabla\cdot\vec{B}=0\]
\[\vec\nabla\times\vec{E}=-\frac{\partial\vec{B}}{\partial{t}}\]
\[\vec\nabla\times\vec{B}=\frac{\partial\vec{E}}{\partial{t}}-g\frac{\partial{f}}{\partial{t}}\vec{E}+g\vec\nabla{f}\times\vec{B}-\Gamma^i_{\mu\beta}\Fub,\]
where 

\begin{equation}
\label{g e f}
g=\frac{\sqrt{1-Ar^2}}{a^3r^2\sin\theta},\mbox{\hspace{2cm}}f=\frac{a^3r^2\sin\theta}{\sqrt{1-Ar^2}},
\end{equation}
and $A=+1,0,-1$. \\
The Wave-equations for \vE~~and \vB~~are found to be given by:

\begin{equation}
\label{equacoes de onda e}
\nabla^2\vec{E}-\ddEt=\na(g\na f\cdot\vec{E})-\dgt\dft\vec{E}-g\ddft\vec{E}-g\dft\dEt+
\end{equation}
\[+\dgt\na{f}\times\vec{B}+g\frac{\partial}{\partial{t}}\na{f}\times\vec{B}+g\na{f}\times\dBt-\frac{\partial}{\partial{t}}(\Gamma^i_{\mu\beta}\Fub);\]

\begin{equation}
\label{equacoes de onda b}
\nabla^2\vec{B}-\ddBt=\na\times(g\dft\vec{E}-g\na{f}\times\vec{B}+\Gamma^i_{\mu\beta}\Fub)
\end{equation}

Now, by virtue of the presence of the gravitational background, we have explicitly built up a solution for \vE~~end \vB~~that evade the claim by Brownstein \cite{K R Browns} and Salingaros \cite{Saling1}.
These authors state that it is always possible to find out parallel solutions 
for \vE~~and \vB~~in Plasma Physics or in an Astrophysic Plasma. However, contrary to their result, we have found non-parallel solutions due to the non-flat background of gravity:
\begin{equation}
\label{Solucoes E}
\vec{E}=\hat{i}(\sin\theta G_{(r,t,\theta)}-Cos\theta F_{(r,t)})kaCos(kz)Cos(\omega t)+
\end{equation}
\[+\hat{j}(\cos\theta F_{(r,t)}-\sin\theta G_{(r,t,\theta)})ka\sin(kz)\cos(\omega t)+\]
\[+\hat{k}[(\sin\theta \cos\varphi F_{(r,t)}+\cos\theta \cos\varphi G_{(r,t,\theta)})ka\cos(kz)\cos(\omega t)+\]
\[-(\sin\theta \sin\varphi F_{(r,t)}+\cos\theta \sin\varphi G_{(r,t,\theta)})ka\sin(kz)\cos(\omega t)]\]
and
\begin{equation}
\label{Solucoes B}
\vec{B}=ka[\hat{i}\sin(kz)+\hat{j}Cos(kz)]Cos(\omega t)
\end{equation}
where the functions $G_{(r,t,\theta)}=\frac{a Cotg\theta}{3\dot{a}r}$ and

\begin{equation}
\label{Solucoes B}
F_{(r,t)}=\frac{2a}{3\dot{a}r}+\frac{Aar}{3\dot{a}(1-Ar^2)}
\end{equation}
are the metric contribution.

\section{Stationary Gravitational Waves and Gravitons}
Now, we analyse the possibility of finding stationary gravitational waves. From a phenomenological viewpoint, a distribuction of black holes could play the role of knots for the non-propagating gravitational waves. We postulate the equation that may lead to this sort of waves:

\begin{equation}
\label{postulado 1}
R_{\mu\nu}=\kappa\Lambda h_{\mu\nu},
\end{equation}

\begin{equation}
\label{postulado 2}
g_{\mu\nu}(x)=\eta_{\mu\nu}+\kappa h_{\mu\nu},
\end{equation}
where $\Lambda$ is the cosmological constant. These equations yield:

\begin{equation}
\label{equacao com aproximacao de campo fraco} 
\partial_\beta\partial_\nu h^\beta_\mu+\partial_\beta\partial_\mu h^\beta_\nu-\Box h_{\mu\nu}-\partial_\mu\partial_\nu h^\beta_\beta=\Lambda h_{\mu\nu}.
\end{equation}
Now, solution of the form

\begin{equation}
\label{solucao da equacao com aproximacao de campo fraco} 
h_{\mu\nu}=C_{\mu\nu}(z)f(t),
\end{equation}

\begin{equation}
\label{matriz solucao da equacao com aproximacao de campo fraco} 
h_{\mu\nu}=\pmatrix{A_{00}&0&0&0\cr0&A_{11}&A_{12}&0\cr0&A_{12}&-A_{11}&0\cr0&0&0&A_{00}\cr}e^{i\tilde{k}z}\cos\omega t,
\end{equation}
can be found, where $A_{00}$, $A_{11}$ and $A_{12}$ are free parameters, whereas ${\tilde{k}=\sqrt{\Lambda-\omega^2}}$ is the wave number. Having in mind that $\Lambda$ is a small number, the frequency $\omega$ must be extremely small. This forces on us to search for a mechanism to detect such low-frequency stationary waves.\\

The equations of motion stemming from \ref{postulado 1} may be derived from the density Lagrangian

\begin{equation}
\label{lagrangiano} 
{\cal{L}}_H=\frac{1}{2}H^{\mu\nu}\Box H_{\mu\nu}-\frac{1}{4}H\Box H-\frac{1}{2}H^{\mu\nu}\partial_\mu\partial_\alpha H^\alpha_\nu-\frac{1}{2}H^{\mu\nu}\partial_\nu\partial_\alpha H^\alpha_\mu+
\end{equation}
\[-\frac{1}{2}\Lambda H^{\mu\nu}H_{\mu\nu}+\frac{1}{4}\Lambda H^2,\]
where
\[H^\alpha_\nu=h^\alpha_\nu-\frac{1}{2}\delta^\alpha_\nu h\] and the bilinear form correspondent operator of Lagrangian \ref{lagrangiano} is given by

\begin{equation}
\label{propagador} 
\Theta_{\mu\nu,\kappa\Lambda}=(\Box-\Lambda)P^{(2)}-\Lambda P_m^{(1)}+\frac{5}{2}(\Box-\Lambda)P_s^{(0)}-\frac{(\Lambda+3\Box)}{2}P_w^{(0)}+
\end{equation}
\[+\frac{\sqrt{3}}{2}(\Lambda-\Box)P_{sw}^{(0)}+\frac{\sqrt{3}}{2}(\Lambda-\Box)P_{ws}^{(0)},\]
and $P^{(i)}$, i=0,1,2, are spin-projetor operators in the space of rank-2 symmetric tensors. The graviton propagator read off from this Lagrangian is given by:

\begin{equation}
\label{funcao theta} 
\Big{<}T(h_{\mu\nu}(x);h_{\kappa\lambda}(y))\Big{>}=i\Theta^{-1}_{\mu\nu,\kappa\lambda}\delta^4(x-y)
\end{equation}
where

\begin{equation}
\label{propagador} 
\Theta^{-1}=\Big{[}XP^{(2)}+YP_{m}^{(1)}+ZP_{s}^{(0)}+WP_{w}^{(0)}+RP_{sw}^{(0)}+SP_{ws}^{(0)}\Big{]}_{\mu\nu,\kappa\lambda}
\end{equation}
with

\begin{equation}
\label{}
X=-\frac{1}{\Lambda-\Box};\qquad Y=-\frac{1}{\Lambda};\qquad Z=-\frac{\Lambda+3\Box}{\Lambda^2+8\Lambda\Box-9\Box^2};
\end{equation}
\[W=-\frac{5}{\Lambda-9\Box};\qquad R=-\frac{\sqrt{3}}{\Lambda+9\Box};\qquad S=-\frac{\sqrt{3}}{\Lambda+9\Box};\]
From this propagator, we can set a current-current amplitude and discuss tree-level unitarity \cite{Carlos 1}. Three massive excitations were found: They are a spin-2 quantum with mass equal to $k^2=\Lambda$ and two massive spin-0 quanta with masses equal to $k^2=\Lambda$ and $k^2=-\frac{1}{9}\Lambda$. The spin-2 is a physical one: the imaginary part of the residue of the amplitude at the pole $k^2=\Lambda$ is positive, so that it does not lead to a ghost. It remains to be shown that the tachyonic pole, $k^2=-\frac{1}{9}\Lambda$, is non-dynamical or decouples through some constraint on the sources. \\

We conclude, then, that in a gravitational background it is always possible to find non-parallel electric and magnetic fields. It is the gravitational field that breaks the parallel configuration of $\vec{E}$ and $\vec{B}$ argued by \cite{K R Browns,Saling1}. Also, a stationary gravitational wave equation was postulated and a particular solution was found. We argue that such a kind of solution is possible to be found in Black Hole distributions. Finaly we set up an effective quantum gravity model where the necessary condition for the
tree-level unitarity for the spin-2 sector is respected. The model is infrared finite though non-renormalizable in the ultravioled limit.\\\\
\noindent{\bf \bf \bf \bf \Large Acknowledgements}\\ 

The authors acknowledge Dr. Berth Schr\"oer for helpful suggestions and technical discussions. Thanks are also due to Gentil O. Pires and
Manoelito M. de Souza for a critical reading of the manuscript. This work was partially supported by the Conselho de Desenvolvimento Cient\'{\i}fico e Tecnol\'ogico - CNPQ - Brazil.

\end{document}